\documentclass[twocolumn,secnumarabic,amssymb, amsmath, nofootinbib,tightenlines,
nobibnotes,showpacs, superscriptaddress, aps, prb]{revtex4}

\usepackage{latexsym}        
\usepackage{amssymb}
\usepackage{amsmath}
\usepackage{amsfonts}
\usepackage[dvips]{graphicx} 

\begin{document}
\title{Coexistence of different electronic phases in the
K$_{0.8}$Fe$_{1.6}$Se$_2$ superconductor: a bulk-sensitive hard
x-rays spectroscopy study.}

\author{L. Simonelli}
%\email{laura.simonelli@esrf.fr}
\affiliation{European Synchrotron Radiation Facility, BP220, F-38043 Grenoble Cedex, France}

\author{N.L. Saini}
\affiliation{Dipartimento di Fisica, Universit{\'a} di Roma ``La Sapienza" - P. le Aldo Moro 2, 00185 Roma, Italy, EU}

\author{M. Moretti Sala}
\affiliation{European Synchrotron Radiation Facility, BP220, F-38043 Grenoble Cedex, France}

\author{Y. Mizuguchi}
\affiliation{National Institute for Materials Science, 1-2-1 Sengen,
Tsukuba 305-0047, Japan and JST-TRIP, 1-2-1 Sengen,Tsukuba 305-0047, Japan}

\author{Y. Takano}
\affiliation{National Institute for Materials Science, 1-2-1 Sengen,
Tsukuba 305-0047, Japan and JST-TRIP, 1-2-1 Sengen,Tsukuba 305-0047, Japan}

\author{H. Takeya}
\affiliation{National Institute for Materials Science, 1-2-1 Sengen,
Tsukuba 305-0047, Japan and JST-TRIP, 1-2-1 Sengen,Tsukuba 305-0047, Japan}

\author{T. Mizokawa}
\affiliation{Department of Physics, University of Tokyo, 5-1-5
Kashiwanoha, Kashiwa, Chiba 277-8561, Japan and Department of Complexity Science and Engineering, University of Tokyo, 5-1-5
Kashiwanoha, Kashiwa, Chiba 277-8561, Japan}

\author{G. Monaco}
\affiliation{European Synchrotron Radiation Facility, BP220, F-38043 Grenoble Cedex, France}

\date{\today}% It is always \today, today,

\begin{abstract}
We have studied electronic and magnetic properties of the
K$_{0.8}$Fe$_{1.6}$Se$_2$ superconductor by x-ray absorption and
emission spectroscopy. Detailed temperature dependent measurements
alongwith a direct comparison with the binary FeSe system have
revealed coexisting electronic phases: a majority phase with high spin
$^{3+}$Fe state and a minority phase with intermediate spin
$^{2+}$Fe state. The effect of high temperature annealing suggests
that the compressed phase with lower spin $^{2+}$Fe state is directly
related with the high T$_{c}$ superconductivity in the title system.
The results clearly underline glassy nature of superconductivity in
the electronically inhomogeneous K$_{0.8}$Fe$_{1.6}$Se$_2$, similar to
the superconductivity in granular phases.
 \end{abstract}


\maketitle

\begin{figure}
	\includegraphics[width=\linewidth]{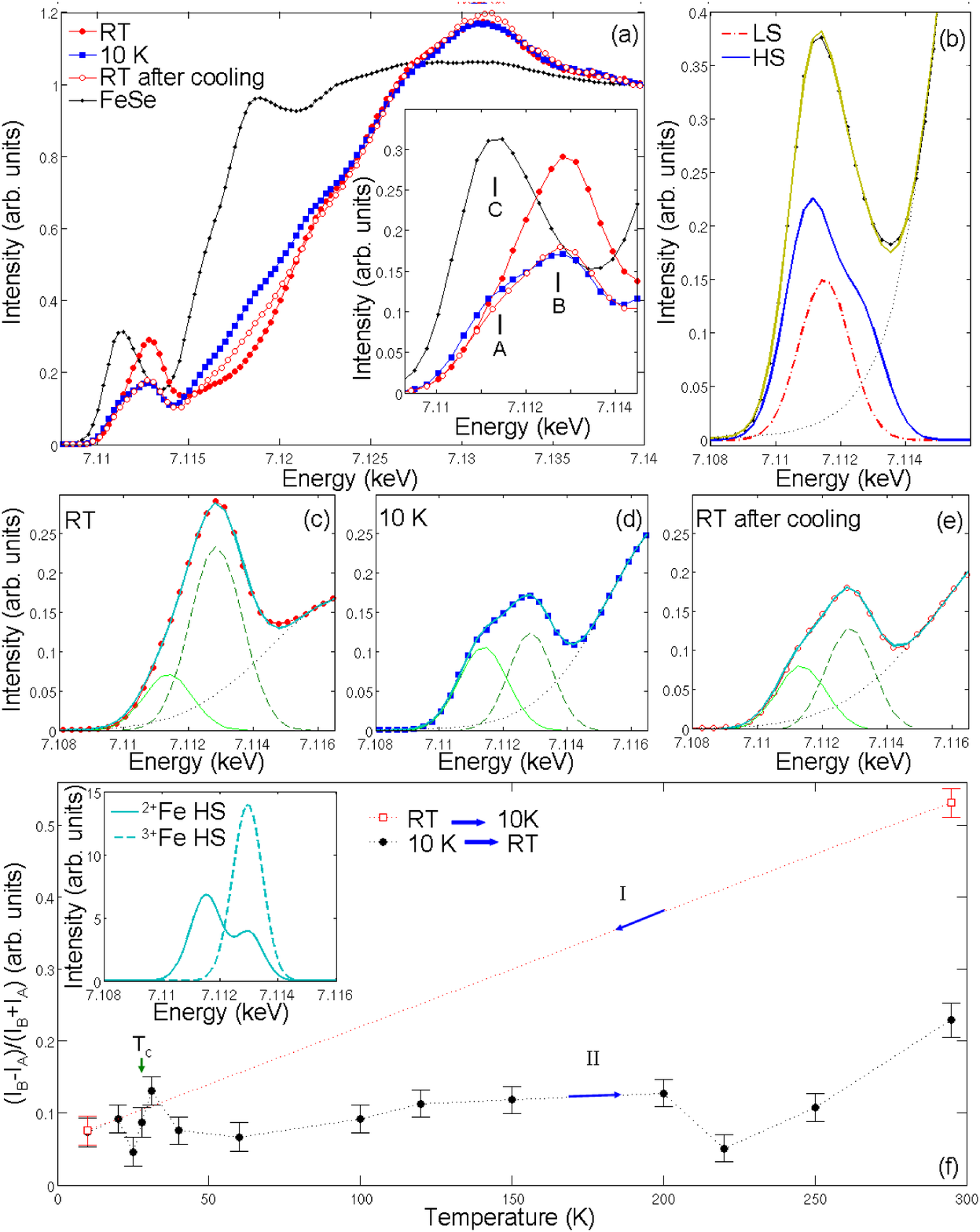}
\caption{(a) Partial fluorescence yield (PFY) spectra measured at the
Fe K-edge of K$_{0.8}$Fe$_{1.6}$Se$_2$ at RT, 10K, and RT after
cooling, compared with the PFY spectrum of binary FeSe.  The inset
shows a zoom over the pre-peak region.  (b) The pre-peak of FeSe is
shown deconvoluted in two components representing the low (dot-dashed
line) and high (solid line) spin $^{2+}$Fe states.  (c) - (e) The
pre-peak of K$_{0.8}$Fe$_{1.6}$Se$_2$ is deconvoluted in two Gaussian
features, A and B. (f) Temperature cycle of the relative intensities
of the two features, (I$_B$-I$_A$)/(I$_B$+I$_A$), while cooling down (I)
followed by warming up to RT again (II).  The inset represents
theoretical simulation of the pre-edge region for tetrahedral high
spin (HS) $^{2+}$Fe (solid line) and $^{3+}$Fe (dashed line)
complexes~\cite{Tami}.}
  \label{F:spectra}
\end{figure}

\section{Introduction}
The discovery of superconductivity in binary FeSe (11-type)
chalcogenide has been an important finding to progress in the
understanding of iron-based superconductors~\cite{Mizuguchi2010}.  The
11-type chalcogenides have been regarded as model systems to explore
the fundamental electronic structure of the iron-based superconductors
since, unlike the more common RFeAsO~(R~=~La,~Nd,~Pr,~Sm,~Gd) and
(Ba,Sr)Fe$_2$As$_2$ pnictides, they lack the spacer layers and hence
the central role played by the Fe-Fe plane with interacting anions
(pnictogen/chalcogen) can be distinctly identified.

Very recently, FeSe layers have been successfully intercalated by
alkaline atoms, with intercalated
A$_x$Fe$_{2-y}$Se$_2$~(A~=~K,~Rb,~Cs) system showing superconductivity
up to 32~K~\cite{Guo,Mizuguchi,Ying,Krzton-Maziopa,Ming-Hu}, unlike
the binary FeSe with a maximum T$_c\sim$~8 K. This new
A$_x$Fe$_{2-y}$Se$_2$ (122)-type superconductor displays a large
magnetic moment per Fe site, intrinsic Fe vacancy order in the
$ab$-plane and an antiferromagnetic order in the
$c$-direction~\cite{Ryan,Bao}.  Several experiments have indicated
that the superconductivity occurs only in Fe-deficient
samples~\cite{Guo2,Wang,Zhang,Yan} and the ordering of Fe vacancies,
important for their electronic and magnetic properties, can be
controlled by heat treatments
~\cite{Wang,Zhang,Ricci,Bao,Han,Svitlyk}.  In addition, the system is
structurally phase separated at the nanoscale containing iron-vacancy
ordered phase with an expanded in-plane lattice and a coexisting
minority phase with compressed in-plane lattice~\cite{Ricci}.
Considering the complexity of the system it is of vital importance to
investigate the coexisting phases for their electronic and magnetic
properties.

In this work, we have exploited bulk-sensitive high energy
spectroscopies to probe the coexisting electronic phases in the
K$_{0.8}$Fe$_{1.6}$Se$_2$ superconductor. We have used x-ray emission
(XES) and high resolution x-ray absorption (XAS) spectroscopy to study
the electronic and magnetic properties as a function of thermal
cycling.  Starting from room temperature (RT), where the system shows
a nanoscale phase separation ~\cite{Ricci}, we have cooled the sample
across the superconducting transition, and the temperature evolution
of the electronic and magnetic properties is monitored while warming
up to RT. Later, we have annealed the sample up to 606 K and quenched
it down to 10 K for a direct comparison between as-grown and annealed
samples.  The results are discussed alongwith the electronic and
magnetic properties of binary FeSe~\cite{Simonelli} to distinctly
identify the effect of K-intercalation on the electronic properties of
these superconductors.

\section{Experimental details}\label{S:Exp}

Measurements of XES and high resolution XAS
were carried out on a well characterized as-grown single crystal of
K$_{0.8}$Fe$_{1.6}$Se$_2$~\cite{Mizuguchi}.  The experiments were
performed at the beamline ID16 of the European Synchrotron Radiation
Facility.  The experimental setup consists of a spectrometer based on
the simultaneous use of a bent analyzer Ge(620) crystal (bending
radius R=1 m) and a pixelated position-sensitive Timepix
detector~\cite{Ponchut} in Rowland circle geometry.  The scattering
plane was horizontal and parallel to the linear polarization vector of
the incident x-rays beam.  The measurements were carried out fixing
the sample surface~(the $ab$-plane of the K$_{0.8}$Fe$_{1.6}$Se$_2$
single crystal)~at~$\sim$~45$^{\circ}$~from the incoming beam
direction and the scattering angle 2$\theta$ at $\sim$ 90$^{\circ}$.
The total energy resolution was about 1.1 eV~full width at
half maximum.  The samples were placed in a cryogenic environment and
the temperature was controlled with an accuracy of~$\pm$1~K.

\begin{figure*}
	\includegraphics[width=\linewidth]{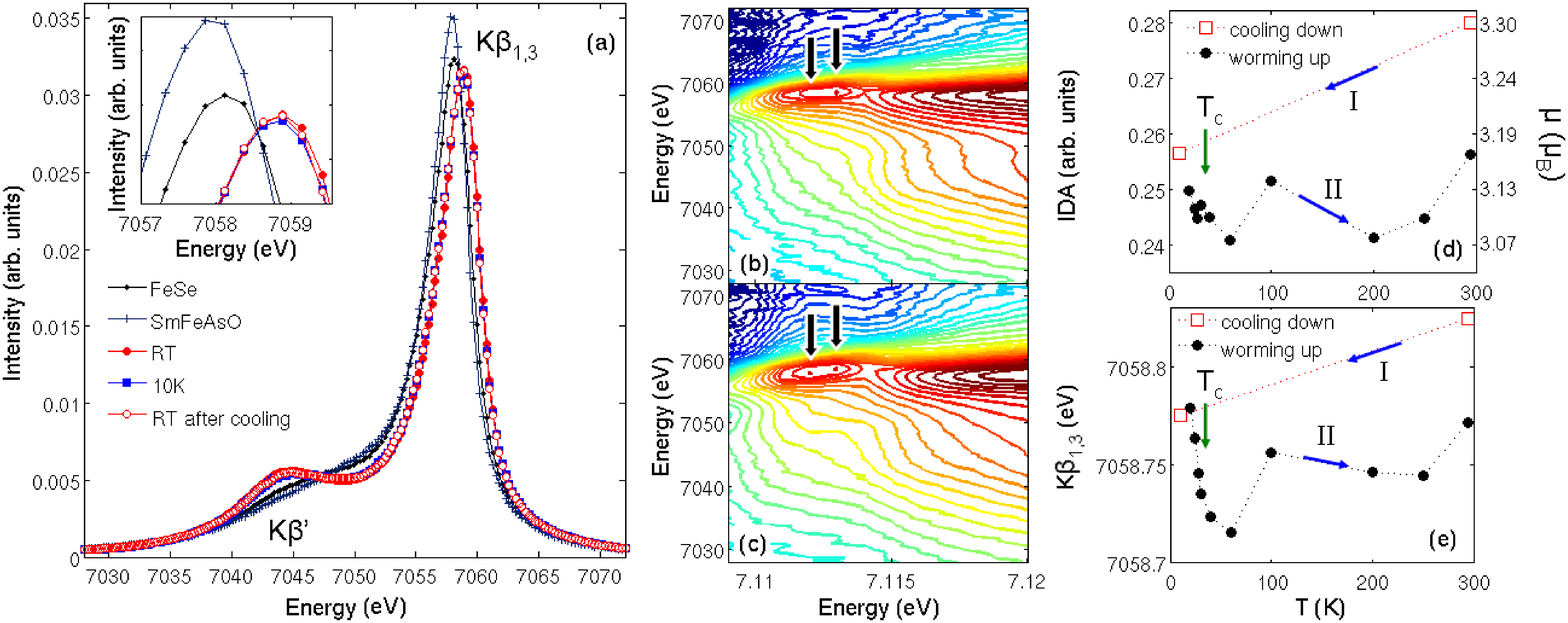}
\caption{(a) Fe K$\beta$ emission spectra of K$_{0.8}$Fe$_{1.6}$Se$_2$
at different temperatures,
compared with the spectra measured on the FeSe and SmFeAsO. The spectra are normalized to the integrated area. The inset
shows a zoom over the main K$\beta_{1,3}$ emission line.  (b)~and~(c)
Resonant x-ray emission maps at RT and 10 K respectively, measured at
the K$\beta$ emission line with the incoming energy around the
absorption pre-peak.  The arrows indicate the two features resonating
at E$_A$ and E$_B$.  (d) IDA as a function of temperature.  Open
squares represent the data while cooling down the sample (I) and the
filled circles represent warming up from 10 K (II).  The scale on the
right hand side shows local magnetic moment ($\mu$) determined
following H. Gretarsson et al.~\cite{Gretarsson}.  (e) Energy position
of the K$\beta_{1,3}$ main line as a function of temperature.  The
uncertainties are of the order of the symbols size.}
  \label{F:spectra}
\end{figure*}

\begin{figure}
	\includegraphics[width=\linewidth]{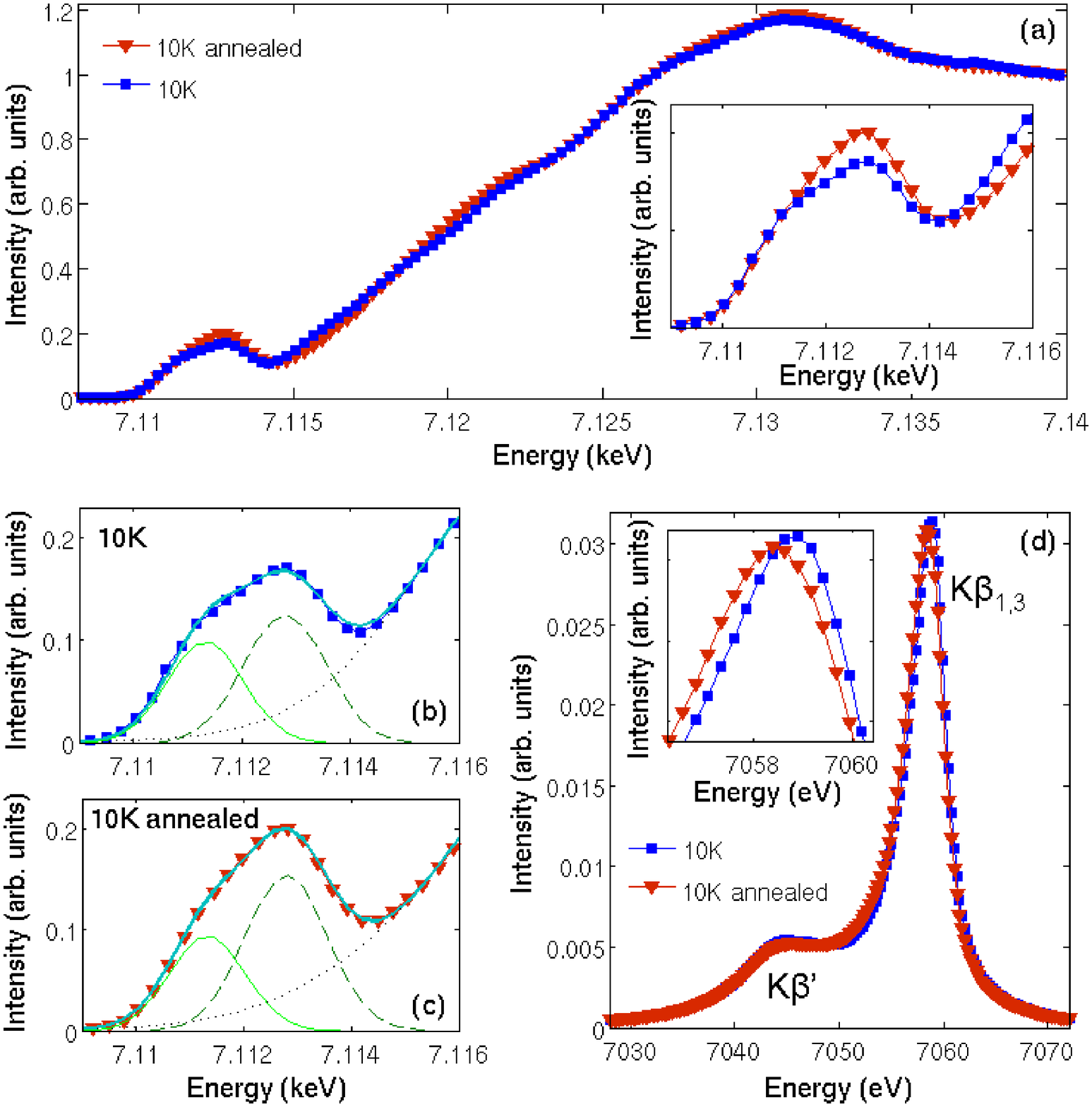}
\caption{(a) Partial fluorescence yield (PFY) spectrum at the Fe
K-edge of K$_{0.8}$Fe$_{1.6}$Se$_2$ measured at 10K after high
temperature annealing (triangles), compared with the PFY spectrum on
as-grown (squares) sample.  The inset shows the zoom over the pre-peak
region.  (b)~and~(c) the pre-peaks before and after annealing are
shown deconvoluted in two Gaussian components.  (d) Comparison of the
Fe K$\beta$ emission spectra collected at 10~K on the sample as-grown
(squares) and after the high temperature annealing (triangles).}
  \label{F:spectra}
\end{figure}

\section{Results and discussions}\label{S:analysis}

Figure~1~(a) shows normalized Fe K-edge partial fluorescence yield
(PFY) absorption spectra of K$_{0.8}$Fe$_{1.6}$Se$_2$, collected at
different temperatures, obtained by collecting the Fe K$\beta_{1,3}$
emission intensity and scanning the incident energy across the
absorption edge.  The spectra are normalized with respect to the
atomic absorption estimated by a linear fit far away from the
absorption edge.  The K-edge absorption process is mainly governed by
the \textit{1$s \rightarrow \epsilon$p} dipole transition.  In
addition, the spectra show strong pre-peak features due to direct 1$s
\rightarrow 3$d quadrupole transitions to the unoccupied Fe \textit{3d}
hybridized with Se \textit{4p} states ~\cite{Joseph}.  In the crystal-field picture the pre-peak is sensitive to
the electronic structure and its energy position, splitting, and
intensity distribution change systematically with spin state,
oxidation state, and local geometry~\cite{Tami}.  The differences
between the FeSe and the K$_{0.8}$Fe$_{1.6}$Se$_2$ are evident.  The
FeSe exhibits a broad and asymmetric single pre-peak feature C at
E$_C$~$\approx$~7111.1 eV (typical of $^{2+}$Fe
complexes~\cite{Chang}) unlike K$_{0.8}$Fe$_{1.6}$Se$_2$, for which
the pre-peak is composed by a doublet feature (see, e.g., the inset of
Fig.~1~(a) showing a zoom over the pre-peak).  The large width and
asymmetry of the FeSe pre-peak feature indicates that this system may
not be a simple low spin $^{2+}$Fe complex, rather it should contain
both low spin (LS) and high spin (HS) $^{2+}$Fe states.  Indeed, as
shown in Fig.~1~(b), the FeSe pre-peak can be deconvoluted in two
components due to LS and HS $^{2+}$Fe states~\cite{Tami}.
FeSe shows then an intermediate spin $^{2+}$Fe state, consistent with
the local Fe-moment of about 2 $\mu_B$ ~\cite{Gretarsson}.  On the
other hand, the PFY spectra of K$_{0.8}$Fe$_{1.6}$Se$_2$ apparently
show a pre-peak with two features A and B, appearing at E$_A$
$\approx$ 7111.4 and E$_B$~$\approx$~7112.9 eV. Since the system
displays a large local Fe-moment of about 3.3
$\mu_B$~\cite{Bao,Yan2,Gretarsson} we expect a doubled absorption
pre-peak to mainly represent the HS $^{2+}$Fe state (however, with the
more intense feature at lower energy as expected
theoretically~\cite{Tami}).  The fact that the spectral weight is
higher at the higher energy, means that HS $^{3+}$Fe state should
coexist with HS $^{2+}$Fe state (see, e.g., inset of Fig.~1~(f)
showing calculated pre-peaks for tetrahedral HS $^{2+}$Fe and
$^{3+}$Fe complexes).  This constructs a direct evidence of coexisting
electronic phases with different spin states in the
K$_{0.8}$Fe$_{1.6}$Se$_2$ system.

Several experiments have revealed an intrinsic phase separation in the
K$_{0.8}$Fe$_{1.6}$Se$_2$ system~\cite{Ricci2,Wang2,Chen} below $\sim$
580 K where iron vacancy-ordered structure is associated to the
magnetic order ~\cite{Bao,Shermadini,Liu}.  Alongwith the majority
phase, a minority phase with slightly compressed in-plane lattice
appears~\cite{Ricci2}.  It is likely that the two phases have
different electronic properties due to local change of the Fe valence
in different sublattices due to the high K mobility and distribution.
In fact, the PFY spectra provide a clear evidence of electronic phase
separation, consistent with the structural nematicity of the system.
The in-plane compressed (out-of plane expanded) minority phase most
likely corresponds to a $^{2+}$Fe oxidation state, while the in-plane
expanded (out-of plane compressed) phase with vacancy ordering to a
$^{3+}$Fe state.  Despite the fact that the phase separation affects
both in-plane and out-of-plane axes, the variation of the $c$-axis is
50$\%$ larger than that along the $a,b$ directions.  Consistently, the
shift of the absorption edge reflects the fact that a $c$-axis
expanded (compressed) phase corresponds to a lower (higher) energy.

In order to investigate the temperature evolution of the electronic
phase separation, we have deconvoluted the pre-peak of
K$_{0.8}$Fe$_{1.6}$Se$_2$ in two Gaussian features, A and B, with fixed
energy positions (Fig.~1~(c)-(e)).  A pseudo-Voigt function was used
to describe the background due to the rising edge (dotted curve).  The
width of A has been kept fixed while that of B was variable to
compensate the effect of the background subtraction.  Figure~1~(f)
displays relative intensity, ($I_B-I_A$)$/$($I_B+I_A$), representing
the qualitative evolution of the coexisting electronic phases.
Starting from RT and cooling down to 10 K, B loses and A gains
intensity.  This may be due to a riorganization between nanoscale
phases while the sample is cooled.  Warming up again to RT,
differently from B, the feature A approximately recovers its initial
intensity, however the distribution of the phases appears different.
The temperature dependence of the relative intensity also reveals a
small discontinuity across T$_c$.  Moreover, a dome like anomaly is
evident between 100 K to 220 K, similar to the resistivity anomaly
seen in this system~\cite{Yan,Fang,Han,Bao2}.
It should be noticed that the Fe-Se bondlength does not shows any change with temperature~\cite{Iadecola12,Tyson}. 
This implies negligible change in the Fe 3$d$ - Se 4$p$ hybridization~\cite{Joseph}. Therefore the absorption pre-peak, consisting 
HS $^{2+}$Fe and HS $^{3+}$Fe components, get redistributed without any appreciable effect of structural changes with temperature.

Independent and complementary information on the electronic and
magnetic properties of the Fe 3d levels can be obtained from XES.  Figure~2~(a) shows Fe K$\beta$ emission
line measured on K$_{0.8}$Fe$_{1.6}$Se$_2$ at different temperatures
alongwith the RT XES spectra of FeSe and SmFeAsO systems.  In the crystal-field picture the overall
spectral shape is dominated by the ($3p$,$3d$) exchange
interactions~\cite{Glatzel}.  In particular, the presence (absence) of
a pronounced feature at lower energy (K$\beta$') is an indication of a
HS (LS) state of Fe~\cite{Glatzel}.  Also, the energy position of the
K$\beta_{1,3}$ provides information on the spin state reflecting the
effective number of unpaired $3d$ electrons~\cite{Glatzel}.
Differences between the XES of K$_{0.8}$Fe$_{1.6}$Se$_2$, FeSe and
SmFeAsO are evident.

In Fig.~2~(b) and (c) we have displayed the resonant x-ray emission
maps collected at the Fe K$\beta$ emission with the incoming
energy around the absorption pre-peak.  The RT map in panel (b) shows
a main feature and a shoulder, resonating at E$_B$ and E$_A$
respectively.  The map measured at 10 K instead (panel (c)) reveals a
clear enhancement of the shoulder intensity, appearing as a distint
feature at a lower emitted energy.  This is a clear indication that
the pre-peak feature A should be related to a lower spin state than
feature B, in agreement with the hypothesis that features A and B
correspond to HS $^{2+}$Fe and $^{3+}$Fe phases respectively.
Therefore, the minority phase has reduced local Fe-moment with respect
to the Fe-vacancy ordered phase.  The observation further underlines
the correlation between the Fe-vacancy order and local Fe-moment in
the title system.

It is possible to quantify the local Fe-moment from the integrated
area of absolute XES difference with respect to a LS
reference~\cite{Gretarsson}.  Since SmFeAsO is almost non-magnetic
~\cite{Gretarsson}, we have taken SmFeAsO as a reference to obtain the
integrated absolute difference (IDA), that is approximately
proportional to the spin magnetic moment.  In order to determine the
relative variation in the magnetic moment, we have used the RT value
of $\mu$ for K$_{0.8}$Fe$_{1.6}$Se$_2$ (3.3
$\mu_B$~\cite{Bao,Yan2,Gretarsson}) and for FeSe (2
$\mu_B$~\cite{Gretarsson}).  Figure~2~(d) and (e) shows temperature
dependence of the IDA and the energy position of the K$\beta_{1,3}$
main line.  Since the two quantities provide the same information,
qualitatively similar trends are observed.  The changes in the K$\beta$ emission are very much consistent with those appearing in the absorption pre-peak, showing a reduced magnetic moment at low temperature due to the redistribution of the phases characterized by the HS $^{2+}$Fe and HS $^{3+}$Fe states. More quantitatively, cooling down, we observe a
reduction of the local Fe-moment from 3.30 to 3.17 $\mu_B$.  Comparing
the corresponding change in the PFY spectra, the reduction of the
Fe-moment is too small if we consider the minority phase being in
low-spin $^{2+}$Fe with no local moment, confirming dominance of
high-spin $^{2+}$Fe state.  In the temperature range from 10 K to 60
K, the $\mu$ continues to decrease down to a minimum of 3.08 $\mu_B$,
unlike a negligible change seen in the PFY. With further increase of
temperature, the local Fe-moment $\mu$ appears to show dome shape
anomalous variation around 150 K, that might be related with the
resistivity anomaly~\cite{Yan,Han,Fang,Bao2}.  This kind of anomalous
behavior can be expected from an inhomogeneous and
glassy system with coexisting phases.

The glassy nature of the K$_{0.8}$Fe$_{1.6}$Se$_2$ is also evident
from the effect of thermal cycling.  To get further insight we have
annealed the sample at 606 K for 20 minutes and quenched it down to 10
K. Figure~3~(a) shows the PFY spectrum after this treatment, compared
with the earlier one at 10 K. There is an apparent change in the
pre-peak with the component B showing a relative increase after the
annealing (Fig.~3~(b) and (c)).  On the other hand, the K$\beta$
emission spectra reveals a clear decrease in the Fe-moment
(Fig.~3~(d)).  Indeed, the IDA for the annealed sample is $\sim$
0.187, that corresponds to the $\mu\sim$ 2.79 $\mu_B$ (to be compared
with $\mu\sim$ 3.17 $\mu_B$ for the as grown sample), more similar to
the moment for the binary FeSe ~\cite{Gretarsson}.  Therefore, the XES
results are consistent with increase of lower spin $^{2+}$Fe state by
annealing.  Apparently this is inconsistent with the PFY spectra in
which the feature B has been assigned to high-spin $^{3+}$Fe state.
However, we should recall that the PFY of FeSe goes through a large
change with pre-peak shifting towards higher energy under hydrostatic
pressure ~\cite{Chen2} while the superconducting transition
temperature changes from a low (8 K) to a high (37 K) value.  Thus an
intensity increase of the feature B in the PFY seems to be due to
compressed low-spin $^{2+}$Fe phase.  Therefore, a distribution of
different phases is likely, with even lower magnetic moment (most
likely related with compressed FeSe filaments in the texture of the
magnetic phase) that could be involved in the high T$_c$
superconductivity of the K$_{0.8}$Fe$_{1.6}$Se$_2$ system.  Indeed,
the system is characterized by a large disorder and glassy local
structure ~\cite{Iadecola12}, similar to the granular materials.

\section{Conclusions}\label{S:conclusions}
In conclusion, we have investigated the evolution of the electronic
and magnetic properties of K$_{0.8}$Fe$_{1.6}$Se$_2$ by XAS and XES
measurements as a function of temperature in a close thermal cycle,
providing a clear evidence of coexisting electronic phases
characterized by different Fe valence and local magnetic moment.
Using resonant XES we have found that the Fe-vacancy ordered
(disordered compressed) phase corresponds to HS $^{3+}$Fe
(intermediate spin $^{2+}$Fe).  The presence of the HS $^{3+}$Fe phase
means that the minority superconducting phase is heavily electron
doped, consistent with the absence of the hole pockets on the Fermi
surface~\cite{ARPES1,ARPES2}.  We also find that the local Fe-moment
sustains a substantial reduction under high temperature annealing due
to disordering of the Fe vacancies.  A comparative study with respect
to the FeSe suggests that high T$_c$ in the title system has a clear
analogy with the increased T$_c$ of binary FeSe under hydrostatic
pressure.  It looks like that the coexistence of different phases is
the key, and since by annealing the disorder increases with a
simultaneous decrease of the local Fe moment, it is likely that a
fraction of different metallic compressed disordered phase in lower
spin configuration ~\cite{Chen} exists.  These filamentary phases get
superconducting as happens in the granular superconductors.

%\begin{acknowledgments}
%\end{acknowledgments}

\end{document}